\documentclass[11pt]{cernrep}
\usepackage{graphicx}
\usepackage{here}
\begin{document}
\title{Similarity renormalization group 
        as a theory of effective particles 
\footnote{ ~~Invited talk at the Light Cone Workshop: 
         {\it Hadrons and beyond}, Durham University, 5-9 August 2003. 
         \hfill {\bf IFT/33/03}}}
\author{Stanis{\l}aw D. G{\l}azek}
\institute{Institute of Theoretical Physics, Warsaw University, Poland}
\maketitle
\begin{abstract}
The similarity renormalization group procedure 
formulated in terms of effective particles is 
briefly reviewed in a series of selected examples
that range from the model matrix estimates of its
numerical accuracy to issues of the Poincar\'e 
symmetry in constituent theories to derivation of 
the Schr\"odinger equation for quarkonia in QCD. 
\end{abstract}
\section{INTRODUCTION}
This talk discusses some examples that illustrate 
recent developments in the effective (constituent) 
particle approach to quantum field theory (QFT). 
The new approach followed in the footsteps and 
emerged from the general renormalization group 
(RG) procedure for Hamiltonians that uses a 
similarity transformation \cite{S} instead of
a cutoff reduction. Complex issues of small-$x$ 
dynamics and chiral symmetry that occur in addition 
to the ultraviolet RG structure in QCD require
extended studies and the effective particle 
picture provides a new framework for them.

\section{ CUTOFF ~$\rightarrow$~ BAND WIDTH $\lambda$}
Canonical Hamiltonians of local QFTs are plagued 
by divergences. The divergences stem from the 
local interactions, which extend over an infinite 
momentum range and involve unlimited numbers of 
bare particles. For example, the term $H_I = \int 
d^3x  \, \bar \psi \hspace{-1.5mm} 
\not \hspace{-1.5mm} A \psi $ produces 
divergent loop integrals. Therefore, one can ask 
what is $H_I$ to mean in the context of  
proper quantum mechanics (QM). For example, if 
already $H_I^2$ does not exist, why should the 
evolution operator $\exp{(-iHt)}$ exist? The left 
drawing on Fig. 1
\begin{figure}[h]
\label{fig:vertex}
\begin{center}
\includegraphics[scale=.6]{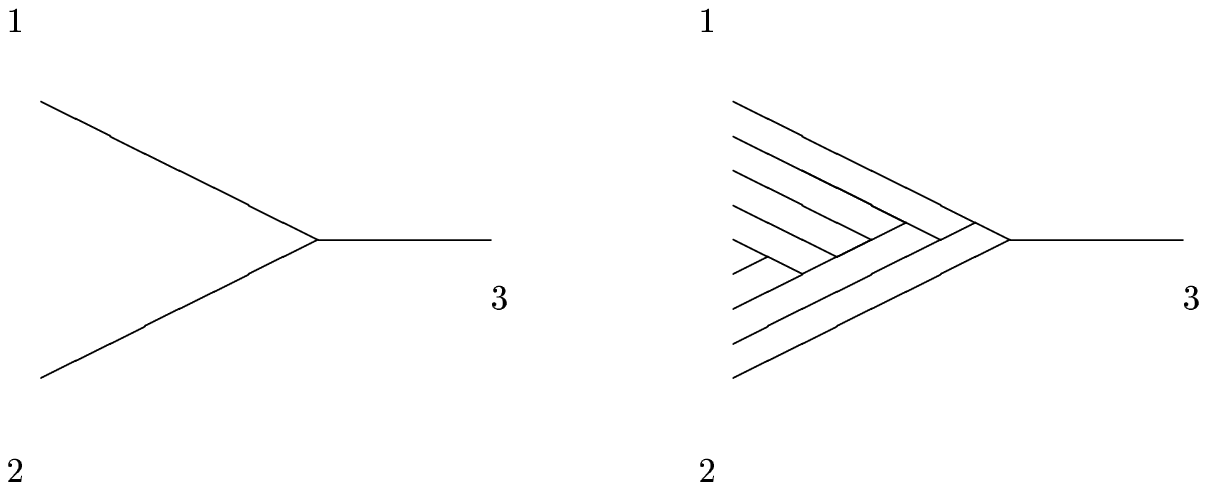}
\caption{ A bare particle decays into two, or many. }
\end{center}
\end{figure}
shows a local interaction that turns one particle into  
two particles whose momenta may be described using standard 
light-front (LF) variables
\begin{eqnarray}
x_1 = k_1^+/k_3^+  \quad, \quad x_2 = k_2^+/k_3^+ \quad , \quad
\kappa^\perp_{12} = x_2 k^\perp_1 - x_2 k^\perp_2 \, .
\end{eqnarray}
Two cutoffs have to be introduced in $H$ in order 
to make powers of $H$ and the eigenvalue problem
for $H$ finite: $|\kappa^\perp|  \le  \Lambda$, and
$x  \ge \delta$. The second cutoff implies that in 
all interactions the number of annihilated or created 
particles is limited by $1/\delta$, since all the $x$s 
must add up to 1 due to translational symmetry. Then, 
the invariant mass $M^2$ of the bare particles in the 
interaction is limited by $M^2 \leq \Delta^2 \sim 
\Lambda^2/\delta$. When the number of particles 
involved in interactions is limited by $1/\delta$, 
it also becomes legitimate to restrict the analysis 
of renormalized theories to polynomial interactions 
in field variables. But the main point is 
that instead of thinking in terms of the bare local 
$H$ itself one needs to think in terms of $H_\Delta$ 
with cutoffs.

Wilson had explained implications of inserting 
cutoffs into Hamiltonians \cite{RG}, and he provided 
a picture of the RG as a family of effective Hamiltonians 
$H_\lambda$, with coupling constants that depend on the 
scale $\lambda$. For example, the counterterms (CT) in 
$H_\Delta$ can change the coupling constant $g$ to $g + 
CT_\Delta(g)$, a cutoff dependent quantity which 
is usually denoted by $g_\Delta$. In the effective 
Hamiltonian with a small running cutoff $\lambda$, a 
similar coupling constant is found, denoted by 
$g_\lambda$. This is illustrated in Fig. 2.
\begin{figure}[h]
\label{fig:wrg}
\begin{center}
\includegraphics[scale=.7]{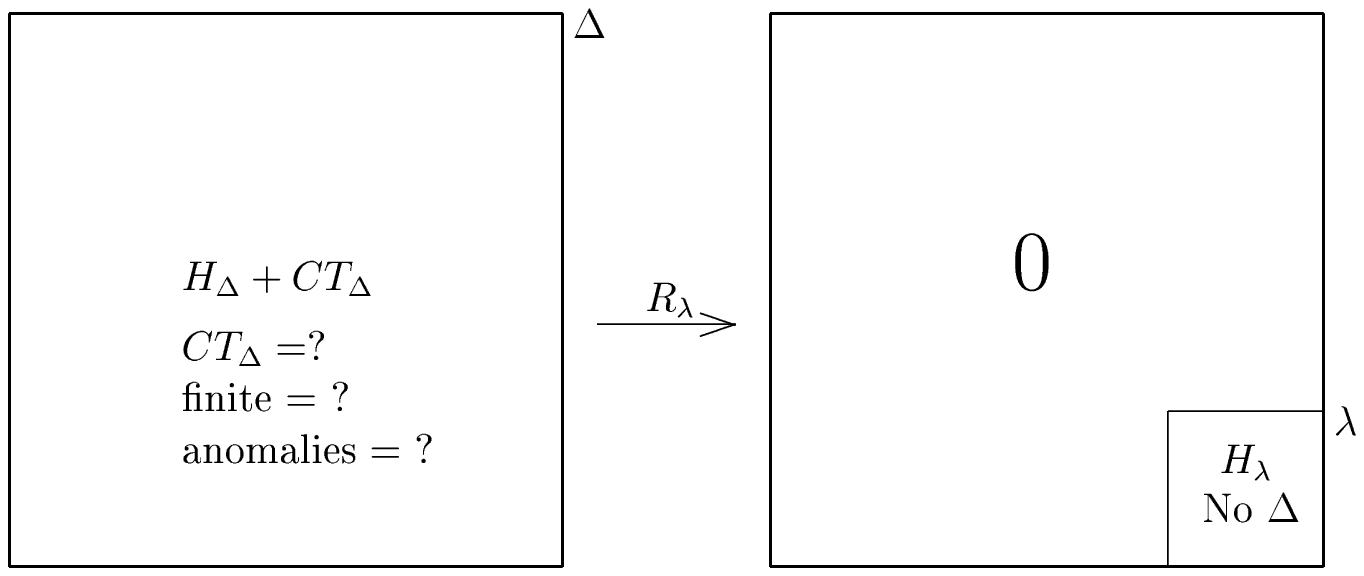}
\caption{ The standard RG procedure with the bare cutoff 
$\Delta$ and the running cutoff $\lambda$. }
\end{center}
\end{figure}
But Wilson's procedure of calculating $H_\lambda$ 
encounters problems with non-trivial interactions 
across the boundary line at $\lambda$ and with small 
energy denominators in perturbation theory at the 
boundary. It becomes very hard to calculate the structure 
of the CTs in asymptotically free (AF) theories with 
bound states (BS) where the overlapping divergences 
cannot be fully controlled unless the BS structure 
is systematically incorporated in the running of the 
coupling constants. Thus, it becomes very hard to 
determine the finite parts of the CTs, control all 
anomalies, and proceed to a meaningful phenomenological 
study using a LF Hamiltonian of QCD. The critical issue
is that one needs a sufficiently small $\lambda$ so that 
the memory and CPU requirements in numerical studies become 
realistic, while small $\lambda$ means that one has 
to reduce the entire theory dynamics to something 
small. This is not easy to do in QCD. But the difficulties 
are partly removed by switching over to the new scheme 
\cite{S} based on the similarity (most often unitary) 
transformation, shown in Fig. 3.
\begin{figure}[h]
\label{fig:srg}
\begin{center}
\includegraphics[scale=.7]{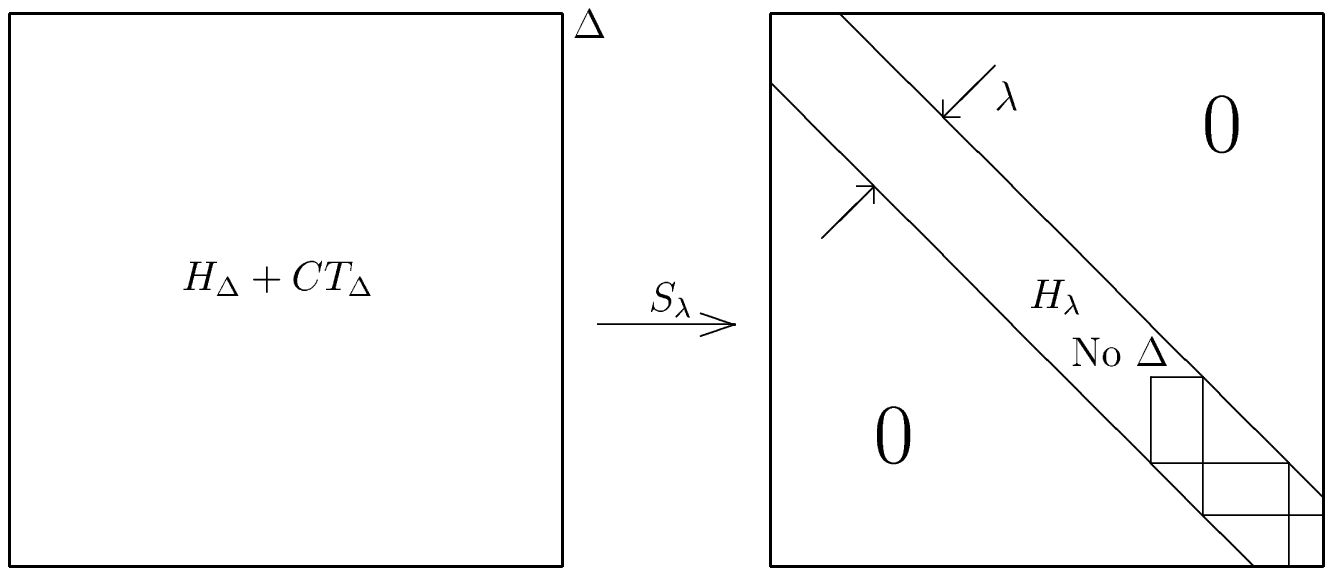}
\end{center}
\caption{ The cutoff $\lambda$ is replaced by the width. 
The zig-zag jumps correspond to perturbation theory (see
the original literature).}
\end{figure}
Beyond this conceptual change, the steps of the RG 
procedure for calculating CTs remain in principle the 
same as in the standard RG approach. Note, however, 
that the genuine infrared divergences due to jumps 
just across the cutoff boundary in Fig.2 are now 
entirely removed from the perturbative evaluation of 
$H_\lambda$ in Fig. 3. The differential equation for 
$H_\lambda$ reads (the initial condition at $\lambda 
= \infty$ is set by $H_\infty = H_\Delta + CT_\Delta$),
\begin{eqnarray}
{d H_\lambda \over d\lambda} = [T_\lambda, H_\lambda] \, .
\end{eqnarray}
$T_\lambda$ denotes the generator that needs to be 
carefully chosen to perform actual calculations.
\section{ PATH TO 1\% ACCURACY }
The idea is to define such $T_\lambda$ that one 
can accurately calculate matrix elements in a 
small window $W_\lambda$ (a sub-matrix of the 
entire Hamiltonian $H_\lambda$, see Fig. 4) 
using perturbation theory, 
\begin{figure}[h]
\label{fig:window}
\begin{center}
\includegraphics[scale=.8]{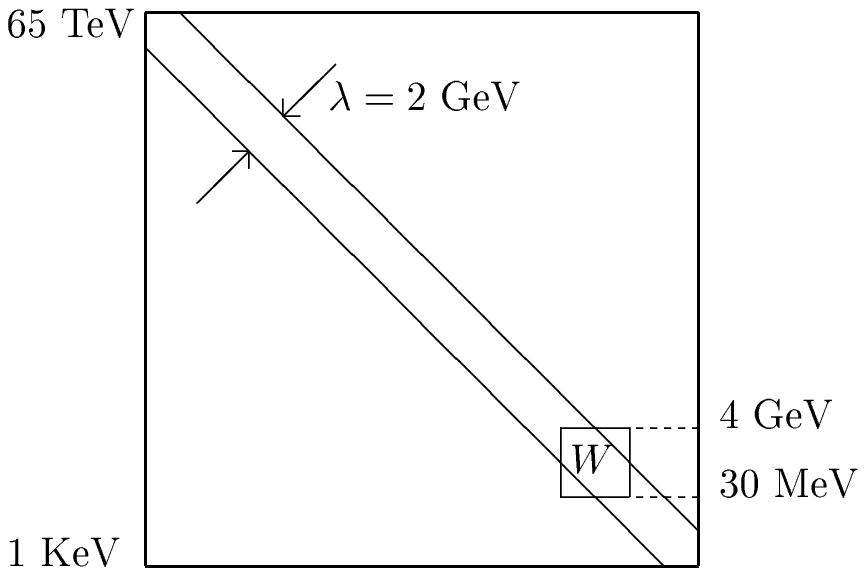}
\caption{ Small window Hamiltonians $W_\lambda$ can be 
diagonalized using computers. }
\end{center}
\end{figure}
and then diagonalize the window $W_\lambda$ 
non-perturbatively to produce the masses and wave 
functions of the BSs in the middle of the energy 
range covered by the window. An elegant candidate 
for the generator can be found in Wegner's flow 
equation \cite{Wegner}: $ T_\lambda = [H_\lambda, 
H_{0 \lambda}]$. A model result obtained using 
this generator \cite{Mlynik} is shown in Fig. 5.
\begin{figure}[h]
\label{fig:wegner}
\begin{center}
\hspace{-1.2cm} 
\includegraphics[angle=270,scale=.25]{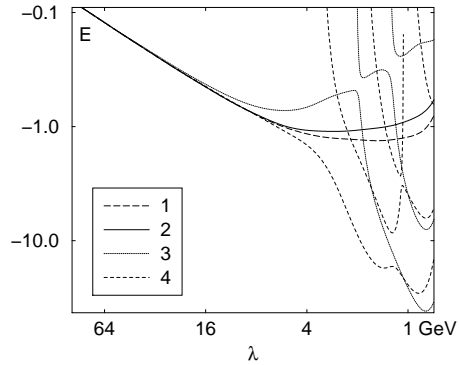}
\caption{ A BS eigenvalue of the window $W_\lambda$. 
The window was calculated in orders $n=1,2,3,4$ of 
perturbation theory in $g_\lambda$ in a matrix model 
of an AF theory using Wegner's generator. The exact 
BS eigenvalue was $E = -1$ GeV.}
\end{center}
\end{figure}
The failure of the theory to converge for $\lambda
< 4$ GeV is clearly visible. But when one uses an
altered version of $T_\lambda$ \cite{Mlynik}, one 
is able to obtain Fig. 6,
which displays evident signs of stable behavior, 
and the accuracy of the procedure reaches 1\%. In 
order to take advantage of this apparently powerful 
method in solving QFT, one can formulate the entire
approach in terms of effective particles.
\begin{figure}[t]
\label{fig:awegner}
\begin{center}
\includegraphics[angle=270,scale=.25]{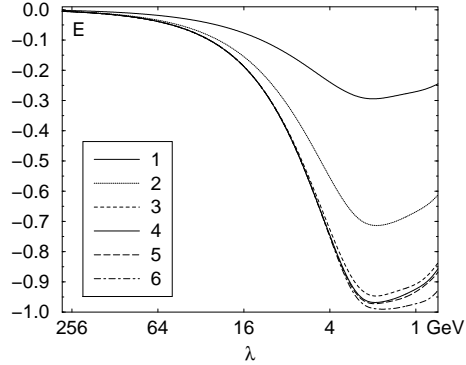}
\caption{ Results from the window $W_\lambda$ in the 
case of the altered Wegner equation, orders 1 to 6, 
cf. Fig. 5. }
\end{center}
\end{figure}
\section{ FROM THE BAND OF WIDTH $\lambda$ TO VERTEX FORM FACTORS $f_\lambda$ }
The RG procedure for effective particles, or RGPEP
(early literature on the subject is referenced in 
\cite{g_QCD}) is based on the observation that 
the initial Hamiltonian $H_\Delta$, expressed in 
terms of bare fields (bare creation and annihilation 
operators, $q$) can be rewritten in terms of products 
of creation and annihilation operators for effective 
particles, $q_\lambda$, with new coefficients, so 
that $H_\Delta(q) + CT_\Delta(q) = H_\lambda(q_\lambda)$, 
and 
\begin{eqnarray}
q_\lambda = U_\lambda q U^\dagger_\lambda \, .
\end{eqnarray}
The calculational procedure is designed so that the 
new coefficient functions in $H_\lambda (q_\lambda)$  
allow us to write $H_\lambda = f_\lambda G_\lambda$, 
where $f_\lambda$ denotes the vertex form factors that 
have momentum-space width $\lambda$, and $G_\lambda$ 
denotes the vertices of the effective interaction 
terms. The generator $T_\lambda$ is chosen so that 
the interactions evolve with $\lambda$ according to 
the equation
\begin{eqnarray}
G'_\lambda = 
[\{(f_\lambda-1)G_\lambda\}', f_\lambda G_\lambda] \, .
\end{eqnarray}
The form factor $f_\lambda$ is a smooth function of 
the effective particle momenta. It secures that the 
Hamiltonian $H_\lambda$ has non-zero matrix elements 
only between states whose total invariant masses 
differ by no more than about $\lambda$. 

Although RGPEP is designed for LF QFT, it can also be 
applied in the matrix model with AF and a BS, which 
was discussed in the previous section \cite{Mlynik}. 
It turns out that the RGPEP equations can produce 
results as good as in Fig. 6, i.e., one {\it can}
calculate the window Hamiltonians $W_\lambda$ with 
accuracy matching the altered Wegner equation. But 
in addition, the RGPEP in LF dynamics offers a boost 
invariant formalism that preserves the cluster property 
of QFT \cite{Weinberg} and introduces a dynamical 
connection between the pointlike partons and effective 
constituent particles of size $\sim 1/\lambda$. 
\section{ RGPEP IN QFT }
Let me observe first that RGPEP explains the 
phenomenon of AF through dependence of 
$H_{LF \,QCD \, \lambda}$ on $\lambda$. For 
example \cite{g_QCD}, the three-gluon coupling 
term ($Y_{123}$ is a polarization dependent factor),
\begin{eqnarray}
H_{I \lambda} & = & \int [123] \, r(x_1)r(x_2) \,
f_\lambda({\cal M}_{12}^2,{\cal M}_3^2)\,\, 
V_\lambda(\kappa^\perp_{12},x_1) \, Y_{123} \,\, 
q^\dagger_{\lambda 1} q^\dagger_{\lambda 2} q_{\lambda 3} \, ,
\end{eqnarray}
defines an effective coupling constant as 
$g_\lambda =  V_\lambda(0^\perp,1/2)$,
which for $f(u,v) = \exp{\left[-(u-v)^2/
\lambda^4\right]}$ and the small-$x$ 
regularization factor $r = x^\delta$ leads in the 
limit $\delta \rightarrow 0$ to the same dependence 
of $g_\lambda$ on $\lambda$ as the one derived for 
the running coupling constant in QCD by Politzer 
\cite{Politzer}, and Gross and Wilczek 
\cite{GrossWilczek} in Feynman diagrams as functions
of the momentum scale. Note also that for other 
small-$x$ regularization factors the effective 
coupling varies with $\lambda$ differently (see 
\cite{g_QCD}). Now, if $g_\lambda$ varies with 
$\lambda$, all physical quantities predictable 
by some window $W_\lambda$ in perturbation theory 
will be expandable in a power series in $g_\lambda$,
too. Thus, the variation of $g_\lambda$ in 
$H_{LF \,QCD \, \lambda}$ provides explanation of 
the asymptotic freedom in observables. On the other 
hand, the structure of counterterms required in the 
Hamiltonian $H_\Delta$ with different regularizations,
and the regularization itself, may be related to 
the worldsheet picture that Thorn discusses in his 
talk \cite{Thorn, BardakciThorn}, and this relation should allow 
us to understand more about the field theory and
strings. One should also mention here that symmetry 
properties of a theory with finite $\lambda$ are 
related to universal RG properties such as coupling 
coherence \cite{Perry}.

Let me then address the issue of the large-momentum 
convergence in dynamics of the effective constituents 
with finite $\lambda$ \cite{Wieckowski}. The usual 
Fock space analysis is imprecise about the notion 
of constituents, there are no form factors $f_\lambda$ 
in the interaction vertices, and the dynamics diverges. 
RGPEP removes the problem by introducing $f_\lambda$. 
One can read about it how $f_\lambda$ works in Yukawa 
theory in a simplest version of the analysis in 
\cite{Wieckowski}. And since the effective constituents 
have the perturbative substructure, the evolution of 
the BS structure functions above the binding scale should 
proceed mainly according to the substructure of the
individual constituents. 

One thus arrives at the major question of a long 
history: is there a Poincar\'e group representation
possible in the picture with a small number of 
constituents? The answer yes is made probable with 
RGPEP according to Ref. \cite{Maslowski}. The trick 
of RGPEP is that the same transformation $U_\lambda$ 
that was used to calculate $H_\lambda$ can also be 
used to transform the entire Poincar\'e algebra 
from the local theory to the one with a small 
$\lambda$. But I have to omit here the details of 
the outline given in \cite{Maslowski}. I limit myself 
only to the indication that RGPEP is a candidate 
for performing Wigner's construction of the 
representations of the Poincar\'e group for 
interacting effective particles in a limited 
range of invariant masses. 

A new and unexpected aspect of the RG studies is
related to the limit cycles that were foreseen
in the past \cite{lc1}, re-discovered in nuclear 
physics recently \cite{lc2}, and now are suggested 
to occur in the infrared domain of QCD \cite{lc3}. 
I want to put on record here that similarity can 
be applied to the mathematical limit cycle model 
studied in \cite{lcmodel}, and as a result one 
obtains a clear numerical example of how limit 
cycles can develop in low energy effective theories, 
associated with formation of BSs.

Let me close this brief enumeration of similarity 
applications in its RGPEP version by returning to 
the issue of quarkonia, since the constituent 
picture in QCD is slowly coming to the foreground 
of studies that involve hadrons \cite{Karliner} and 
the old question of how a theory as complex as QCD 
could be approximated by a constituent model in the 
rest frame of a hadron \cite{Wilsonetal} should also 
be asked within RGPEP. The answer for heavy quarkonia 
has a lot in common with the one developed earlier in 
LF QCD by Perry and collaborators \cite{Brisudova}, 
though one should be aware of important differences. 
According to \cite{QQ}, one can 
derive a Hamiltonian for a pair of effective heavy 
quarks, $H_{Q\bar Q}$, and write its eigenvalue 
equation with the eigenvalue $E=2m+B$ (note the $+$ 
sign), when one assumes that the energy of a gluon in 
the vicinity of a pair of quarks can be described using
some form of effective mass. If the gluon mass is zero, 
one obtains for the quarkonium a Coulomb problem with 
factor 4/3. The interesting result of RGPEP is that in 
the NR limit for very heavy quarks the details of the 
gluon mass ansatz do not count much and as long 
as the mass is sizable the effective $Q\bar Q$ 
eigenvalue equation takes the form (BF denotes the 
Breit-Fermi terms associated with the Coulomb potential)
\begin{eqnarray}
\label{result}
\left[2m - {\Delta_r \over m} 
- {4 \alpha\over 3} \left({1\over r} + BF\right) 
+  {k\over 2} \, r^2 \, \right]\, 
\psi(\vec r) = M \, \psi(\vec r) \, ,
\end{eqnarray} 
where 
$k = m\omega^2/2$ , and $\omega =
\sqrt{ {4\over 3} \, {\alpha \over \pi} } \, 
\lambda \, \left( \lambda \over m \right)^2 \,
(\pi /1152)^{1/4}$, which is in the ball park
of phenomenological scales for reasonable
choices of the parameters. The fascinating aspect of this
result is that the quarks bind above threshold ($B > 0$)
due to the positive self-interactions of quarks,
which constantly emit and re-absorb effective 
gluons in QCD.
\section{CONCLUSION}
In summary, the RGPEP features a single formulation 
of an entire theory (no extra prescriptions for loops 
are needed) in Hamiltonian QM. And the method of
similarity and solving window dynamics to reduce a 
complex theory to a simpler one that uses effective 
degrees of freedom may find application not only in 
particle physics, but also in nuclear, atomic, and 
condensed matter physics, and in chemistry. 

In LF QCD itself, in the effective particle basis in 
the Fock space, one always works with the Minkowski 
metric (no reference to Euclidean geometry is invoked;
and gravity is ignored because it is weak and because 
the quantum nature of masses of effective particles 
is not fully understood) and one keeps intact the boost 
invariance and cluster property in the interactions of 
the effective particles. But the latter have complex 
substructure. Such framework is highly desired for 
theoretical explanation of the parton and constituent 
models of hadrons in QCD. And since the elementary model 
studies reach the accuracy of 1\%, the scheme is 
straightforward, and for heavy quarkonia it can explain 
the binding mechanism above threshold, the question I 
should answer now is what is preventing an immediate 
application to nucleons? My answer is that in order to 
discuss the nucleons one needs a more detailed understanding 
of chiral symmetry, and pions. Although the principles 
of how to deal with chiral symmetry in LF QCD are already 
known \cite{Wilsonetal}, the required quantitative mechanism 
in the Fock space is still a mystery. The possibility of 
a nearby infrared limit cycle \cite{lc3} is certainly 
not making the matter simpler than it was before. The 
current idea is to tackle the 4th order RGPEP in QCD, 
and keep testing the RGPEP approach in QED, since so far 
very little is known about gauge theories formulated in 
terms of the effective particles.
\section*{ACKNOWLEDGMENTS}
I would like to thank C. Thorn for a very interesting 
discussion and S. Dalley for inviting me to this 
stimulating meeting. This work was supported by KBN 
grant number M5/E-343/S.

\end{document}